\documentstyle[preprint,eqsecnum,aps]{revtex}
\begin{document}
\draft
\title
{Decay of the Sinai Well in $D$ dimensions.}
\author{  A.J.Fendrik and M.J.S\'{a}nchez }
\address
{Departamento de F\'{\i}sica, Facultad de Ciencias Exactas y Naturales,\\ 
Universidad de Buenos Aires, Ciudad Universitaria, 1428, Buenos Aires, 
Argentina.} 
\date{\today}
\maketitle
\begin{abstract}

We study the decay law of the Sinai Well in $D$ dimensions and relate the
 behavior of the decay law to internal distributions 
that characterize the dynamics of the system.
We show that  the long time tail of the
decay is algebraic ($1/t$), irrespective of the dimension $D$. 
\end{abstract}
\narrowtext

\section{Introduction}
\label{sec:int}
In a previous work \cite{frs} we studied the decay of quasibounded 
classical 
hamiltonian systems in two dimensions, in particular the
decay problem for  the Sinai well. In this report we extend the study 
to the $D$-dimensional case. We will briefly describe what a quasibounded
system means, but for further details we refer to the original work.\\  
 A quasibounded system is a dynamical system  
transiently bounded to a finite region of the phase space where an infinite 
set of non stable periodic orbits is included before it displays 
unbounded dynamics. The transition from the bounded motion
to the unbounded one is  the decay process, and the decay law is related
to the bounded transient dynamics.
The kind of systems that we are interested in is fully chaotic but not
completely hyperbolic. In terms of the invariant set it means that
 instead of being completely hyperbolic, it has a non hyperbolic (namely 
parabolic \cite{Berry}) subset. 
One of the main differences with the analogous system in two dimensions
is that whereas in that case the invariant set could be fully hyperbolic or 
have a parabolic subset depending on the value of a simple parameter, for the 
$D>2$ dimensional system the invariant set has always a parabolic subset.
In this case the global consequence on the decay law is that it always 
exhibits  a crossover  between a stretched exponential and an algebraic decay
for long times.
One of the main purposes of this work is to study the long time tail of the
 decay law in  $D$ dimensions and to 
relate the decay of population from equilibrium to internal distributions 
that characterize the dynamics.\\
Our work is organized as follows. In Sec.\ \ref{sec:pres} we 
introduce the 
system that we will call the Sinai well in $D$ dimensions because
it is  geometrically similar to the Sinai billiard \cite{sinai} in $D$
dimensions but with  a finite rather than infinite well. 
In Sec.\ \ref{sec:num} 
we review some results of the analogous  model in two dimensions \cite{frs}.\\
Sec.\ \ref{law} is the central body of the work and is devoted to the study 
 of the temporal decay law both numerical and analytically.
We show the results of the numerical study of the decay 
that reveal the above mentioned behavior, and using 
ergodic properties we relate the decay law to internal distributions
that depend on the internal dynamics.
Finally, Sec.\ \ref{con} is devoted to 
discussions and conclusions.
We include Appendix A in which the explicit
expression for the transition probability from the bounded to the free
region is derived.

\section{THE SINAI WELL IN D DIMENSIONS}
\label{sec:pres}
 
Our system consists of a point particle of unit mass in a $D$ ($D>2$)
dimensional 
square well (hypercube) of depth $-V_{0}, (V_{0}>0)$ and side $a$,
 which collides
elastically with a fixed scatterer located in the center of the well.
The scatterer is a $D$ dimensional unit sphere and  as usual we take  
the speed of the particle to be one.
In the study of the analogous system for $D=2$ \cite{frs} we consider 
the total energy $E=p^{2}/2-V_{0}$ $(0\leq E \leq V_{0})$, and explain
in detail how the collision with the  scatterer rearranges the 
energy among the degrees of
 freedom in order that the particle could transit from the bounded region 
to the free one after colliding with the central barrier.\\
Although the extension to higher dimensions is straightforward, there are 
some new features to remark on and we devote the remainder of 
this section to these results.\\
The bounded motion in $D$ dimensions is caracterized by the condition

\begin{equation}
\label{lim}
\sqrt{1- \biggl( \frac{\vec{p} \cdot \hat{n_{i}}}{\mid \vec{p} \mid} 
\biggr) ^{2}} > \sin {\Psi_{lim}} \;, 
\end{equation}
where $\hat{n_{i}}$, $i=1,..,D$ , is the normal direction to face $i$ on 
which the particle bounces and 
 
\begin{equation}
\Psi_{lim} = \arcsin {\frac{1}{\sqrt{1+ V_{0}/E}}} \;, \label{limpr}
\end{equation}
is the limit angle in $D$ dimensions.
 The inequality (\ref{lim}) is the condition to have an 
internal
reflection when the particle reaches the  boundary of the hypercube.
 This result has been explained in detail
for $D=2$ in Appendix A of \cite{frs}, and for $D>2$ the derivation is 
similar. 
As we mentioned before, the collision with the scatterer changes 
the value of the 
components of the momentum and this enables  the transition to the free
region,
or in other words the decay process.
In the $D$ dimensional problem the limit angle $\Psi_{lim}$ can be related 
to the probability $\omega_{D} <1$ 
that the particle transits from the bounded  to the free region after
one collision with the scatterer. The space of momenta is a $D$ dimensional
 unit sphere in which 
\begin{equation}   
\omega_{D} = \frac{d\Psi_{lim}(D)}{\Omega(D)} \:, \label{prob}
\end{equation}
where $d\Psi_{lim}(D)$ and  $\Omega(D)$ are respectively the solid angle 
subtended by $\Psi_{lim}$ and  the total solid angle in $D$ dimensions.
In Appendix A we derive an explicit expression for (\ref{prob}) in 
terms of 
$\Psi_{lim}$. Here we give the result 

\begin{equation}
\omega_{D} \sim \Psi_{lim}^{D-1}
\end{equation}

which shows that the transition probability decreases with the dimensions
$D$ when $\Psi_{lim} \ll 1$ for a given (fixed) energy $E$. 

\section{PRELIMINARY REMARKS}
\label{sec:num}

In this subsection we review some results of  the
analogous problem in two dimensions \cite{frs}.
The decay of population $N(t)$ inside the well
 is characterized  by two distributions specific to the internal 
dynamics. The first 
one  {\it g(t)dt} is the fraction of particles for which the
first collision with the circular scatterer occurs between {\it t} and 
{\it t+dt}, and the second one {\it f(t)dt} is the fraction of particles 
for which the
time between two successive collisions with the central scatterer lies  
between {\it t} and {\it t+dt}. In the decay process from the equilibrium
population these distributions are not independent, but related by

\begin{equation}
dg/dt=-g(0)f(t) \;,
\end{equation}

In  \cite{frs} (in the following I) we conclude 
that distributions  {\it g(t)dt} that 
decrease 
exponentially or faster lead to an exponential decay law while an algebraic
decay of {\it g(t)} gives rise to an exponential decay law that changes 
into a power law decay for long times.
Our model is closely related to the periodic Lorentz gas in two 
dimensions, in which the problem of the asymptotic behavior of the velocity 
autocorrelation function has been studied extensively over the last 
several years
both theoretically and numerically (see \cite{bleher} and references 
cited there). In that model we can distinguish two kinds of behaviors 
depending on
the ``finite'' or ``infinite''nature of the horizon. By definition a periodic 
configuration
of scatterers has infinite horizon if the length of the free motion of the 
particle is unbounded. Actually if the horizon is infinite then there exist
trajectories that never reflect from  the scatterers. These trajectories 
define
the so called corridors that are characterized by the directions of the 
velocity such that
 $v_{y}/v_{x}=z_{1}/z_{2}$ (here $z_{1}$ and $z_{2}$ are 
coprime integer numbers).
The number of open corridors increases when the radii of the scatterers $R$
decrease. Following the notation of ref. \cite{geis} 
 for the square Lorentz model there are at least two of such open channels 
corresponding to x-direction and y-direction, and  we  call these 
channels $\alpha$ and $\beta$ 
respectively. When $\sqrt{5}/10 < R < \sqrt{2}/4$ we have 
other open channels ($\gamma$) that correspond to $v_{y}/v_{x}=\pm 1$.
In I we explained in detail the connection between
 the open corridors and the parabolic periodic orbits  that appear
in our system when the radius of the central scatterer decreases.
For the Sinai well a finite horizon, which means {\it g(t)dt} 
decreasing exponentially or faster, is compatible with the existence
 of periodic orbits 
of hyperbolic type (with no open corridor in the extended Lorentz version),
 whereas an infinite horizon, {\it g(t)dt}
with algebraic tail, implies the existence of 
parabolic non-isolated periodic orbits that appear for $R\le R_{c_{i}}$ and
 determine the corresponding corridors.\\
In the former case the decay law is exponential, and in the latter the decay
 law shows a crossover between an
exponential and a power law decay ($ \sim 1/t$) for long times.\\

\section{THE DECAY LAW IN D DIMENSIONS}
\label{law}
\subsection{NUMERICAL STUDY OF THE DECAY}
In this subsection we show the results of the numerical  study of the decay.
It is a well known fact that, for short times, the behavior of the decay 
law in $D$ dimensions is of exponential type \cite{douss}.\\
Our main purpose is to understand the behavior for long times. In other 
words, we study the long time tail of the decay law in $D$ dimensions.
As we will see in the following section, in order to extract
information about the long time tail of the decay it is enough to study the
function {\it g(t)dt} , the fraction of initial conditions for which the 
first collision
with the scatterer occurs between $t$ and $t+dt$, 
for the $D$ dimensional system 
(to be more precise we will compute the integral of $g(t)$, 
 $G= \int_{t'=0}^{t'=t} g(t')  \;  dt'$ ).
{}From the numerical point of view, the study of $G$ as a function of $t$ 
instead of the  decay law $N(t)/N_{0}$ 
has a great advantage because it requires less CPU time and we can get
 better statistics with more initial conditions. 
We begin with $N_{0}=10^{7}$ particles
with random initial conditions in the phase space and the 
ratio $V_{0}/E=20$.
Fig.~\ref{234gr2p3}(a) shows the results of the numerical study of $G$ for
 $D= 2,3,4$ and radius of the $D$ dimensional scatterer $R=0.23$. 
The behavior is exponential 
for short times and becomes algebraic for long times $(1/t^{\delta})$.  
Fig.~\ref{234gr2p3}(b) shows the tails of the $G$ for $D= 2,3,4$ 
 together with the best fit that predicts for all the curves an exponent
$\delta= 1$.
Fig.~\ref{234gr4p0}(a) is the same as Fig.~\ref{234gr2p3}(a) but for $R=0.4$.
For $D=2$ the $G$ is of exponential type.
This agrees with the result of  paper I. For the two-dimensional
system $R>R_{c1}= \sqrt{2}/4$ is compatible with a $g(t)$ of 
"finite" horizon and the decay is exponential for all times.
For $D=3,4$ again the best fit of the long time tail predicts an exponent
 $\delta= 1$. Fig.~\ref{234gr4p0}(b) shows this fit together with the
 numerical results of the long time tails for $D= 3,4$. 
These results suggest that for $D>2$ there does not exist a critical 
value of the radius that changes the behavior of the decay for long times
as  happens for $D=2$ (see I). In terms of the invariant trapped set, we can 
stress that
 for $D>2$ there are always periodic orbits of parabolic type, and the
initial conditions asymptotic to them are the ones that contribute 
to the long time
tail of the decay.
 
\subsection{THEORETICAL STUDY OF THE DECAY}
One of the main results of the present paper  is
 that for the Sinai well in $D$ dimensions the long time tail of
the decay law is $( \sim 1/t)$, independent of the dimension $D$.
To our knowledge this is the first report in which this result is 
stated analytically and confirmed by  numerical simulations. All the previous 
works only study the model in two dimensions \cite{scheu}, \cite{bin}
or conjecture for the $D>2$ system an exponential decay of the velocity 
autocorrelation function for long times \cite{douss}.
To begin  the theoretical study,
we extend to the $D$ dimensional case some results of the previous paper I
 that allow us to relate the decay law to the internal dynamics: 

\begin{equation}
\hat{Q}(s)= \frac {\omega_{D} \; \hat{g}(s)/s}{1-(1-\omega_{D})\hat{f}(s)} \;,
\end{equation}
where $\hat{Q}(s)={\it L}[Q(t)]$ means the Laplace Transform and
\begin{equation}
\label{qt}
Q(t)= 1- \frac{N(t)}{N_{0}}
\end{equation}
and $N(t)/N_{0}$ is the fraction of particles present in the well at time 
$t$ (the decay law).
Taking into account the equation, 
\begin{equation}
dg/dt=-g(0)f(t) \;,
\end{equation}
so that 
\begin{equation}
\hat{f}(s)=1-s\hat{g}(s)/g(0) \;,
\end{equation}
we finally  find
\begin{equation}
\hat{Q}(s)= \frac{\omega_{D} \; \hat{g}(s)/s}{1+(1-\omega_{D})
[\hat{g}(s)s/g(0) -1]}\;. \label{conv}
\end{equation}
The preceding equation is a straightforward extension of the one that we 
have obtained in I for the two-dimensional problem, being in that case
 $\omega_{D=2}= w$.
To know $Q(t)$, we must be able to inverse-transform (\ref{conv}).
Computing the leading term of (\ref{conv}) and employing (\ref{qt}) we 
obtain
\begin{equation}
\label{tail}
N(t) \sim \omega_{D} \int_{t'=0}^{t'=t} g(t') dt' 
\end{equation}
We must center our attention on the function {\it g(t)dt}
for our $D$ dimensional problem to study the long time behavior
(\ref{tail}) of the decay law.

We will extend the definition of corridors to higher dimensions.
For that we must appeal to the Periodic Lorentz Gas in the $D$-dimensional
case which statistical properties in the hyperbolic domain (finite horizon)
have been studied in detail in a recent reference \cite{cher} and a numerical 
study of some universal properties was performed in \cite{douss}.
We will say that a $D$-dimensional 
Periodic Lorentz Gas has infinite horizon if the length of the free motion is 
unbounded. 
The first trivial extension is to define the corridors in $D$ 
dimensions by
the 
directions of the velocity $\vec{v}$ that satisfy $v_{i} / v_{j}=z_{1} /
z_{2}$
$\forall (i,j)$, where $z_{1}$ and $z_{2}$ are coprimes integers.
These are not the only directions that lead in $D$ dimensions to unbounded 
free
motion. For example for $D=3$ the direction $v_{1}=0$, 
$v_{2}$ and $v_{3}$ arbitraries (compatible with the condition that the 
modulus
of the velocity, $\mid \vec{v} \mid$, is one) is also a corridor in the 
meaning of free unbounded motion.
The main characteristic to note is that as the dimension increases it 
is possible to find 
ever more directions that define unbounded motion  for any value of the 
radius of the scatterer $R \le{0.5}$ which means that there is no  critical 
value $R_{c}$ that destroys the corridors, as there was in two dimensions. 
However $R_{c}$ influences on the number of open corridors, 
since there are some corridors that disappear when $R>R_{c}$.\\  
All the trajectories that contribute to the long time tail of the decay
lie almost entirely in some corridor. In terms
of the Sinai Well we say that these initial conditions  are 
asymptotic to the parabolic periodic orbits that exist in $D$ dimensions, 
leaving the bounded region with probability $\omega_{D}$ after
one collision with the $D$ dimensional scatterer.\\
At this point it is necessary to emphasize that apart from the corridors 
which asymptotic initial conditions contribute to the algebraic long time tail
of the decay with an exponent $\delta=1$ (we call them principal corridors),
there are other corridors in which initial conditions
asymptotic to them lead to an algebraic  tail of the decay 
$(1/t^{\mu})$ with the exponent $\mu>1$. 
A proof of the existence
 of these other corridors is given in Fig.~\ref{4gr4p0st}. It shows for $D=4$ 
and $R=0.4$ two curves $G$ as a function of $t$.   
In one (dashed line) all the corridors have been populated 
and the exponent  for the long time tail is $\delta= 1$, 
in the other (solid line) we  do not consider  initial conditions 
asymptotic to the principal corridors, resulting in an  exponent for the long 
time tail different from $1$.\\
Since the global behavior  of
the decay law for long times is $(1/t)$ independent of the dimension $D$,
we call these other corridors hidden corridors, because there is no evidence
of them in the long time tail of the decay law. 
On the other hand the intermediate behavior of the decay law,
that it is related to asymptotic conditions to the hidden corridors,  
results in a superposition of algebraic decays  with exponents greater than
 and 
different from one, this behavior being more and more complicated as the 
dimension increases.\\
We devote the remainder of this section to derive the explicit dependence
on $t$ of the distribution {\it g(t)dt} for initial conditions that are 
asymptotic to the principal corridors using the $D$-dimensional 
Periodic Lorentz Gas model.
Without loss of generality we will compute {\it g(t)dt} for the initial
conditions that lie in the principal corridor ($D=3$) defined 
by the directions of
 the velocity  $v_{1}/v_{2}=\pm 1$ and $v_{3}$ arbitrary that satisfy the
condition $\mid \vec{v}\mid =1$, corresponding this case to $R<R_{c}$. 
In fig.~\ref{config}  we show a schematic representation of the mentioned
corridor of width $l$ in which we have cut with the plane $z=const.$ 
the spherical
 scatterers at their centers in order to simplify the figure.\\
 We remark that as  $\mid\vec{v}\mid=1$ the distributions in times are 
equivalent to distributions in lengths.\\   
Let $n(t^{*})$ be the fraction of initial conditions for which the first 
collision
 with some scatterer occurs in times $t>t^{*}$.
 As we are interested in the large $t$ behavior the angle $\alpha$ is 
proportional to $n(t^{*})$
\begin{equation}
n(t^{*})\sim \alpha \sim \frac{l}{t^{*}} 
\end{equation}
and
\begin{equation}
-dn/dt^{*} \sim \frac{l}{t^{*2}} \sim g(t^{*})
\end{equation}

where ${\it g(t^{*})dt^{*}}$ is the fraction of initial conditions for
which the first collision with the scatterer occurs between $t^{*}$ 
and $t^{*}+dt^{*}$. To be more precise we must consider the integration
over the  solid angle, but this gives $2\pi$, so we can conclude that 
inside a principal corridor

\begin{equation}
g(t) \sim \frac{cte}{t^{2}} .
\end{equation}

Use of  the last expression  in (\ref{tail}) leads to the mentioned
behavior of the long time tail of the decay 
\begin{equation}
N(t) \sim \frac{\omega_{D}}{t} \; .
\end{equation}  

We stress that the last derivation is also valid for the other principal 
corridors. In the case of the hidden corridors, vestiges of some principal 
corridors when $R>R_{c}$, the integration over the solid angle leads
to an additional $1/t$ dependence, so in that case
\begin{equation}
g(t) \sim \frac{cte^{'}}{t^{3}} \; .
\end{equation}
 
The preceding arguments can be extended to higher dimensions. For the
principal
corridors the dependence on $t$ is exactly the same $(\sim 1/t)$. For the
hidden corridors as the dimension $D$ increases the integration over the
 solid angle gives rise to contributions of the type $1/t^{\mu}$ with 
the $2\leq\mu\leq{D}$.

 \subsection{The Decay Law in terms of the Laplace Transform}

We begin computing  the decay law $N(t)$ as the Inverse Laplace 
Transform of a function $q(\lambda)$, 
\begin{equation}
\label{ilt}
N(t) = \int_{^0}^{\infty} q(\lambda) \;  e^{-\lambda t} \; d\lambda
\end{equation} 
where $q(\lambda)$  can be seen as the fraction of initial 
conditions that decay at time $t$ and which decay rate 
varies between $\lambda$ and $\lambda +d\lambda$.\\
In the context of (\ref{ilt}) the whole decay law 
can be thought as the result of infinitely many decay processes of exponential

type each one characterized by a decay rate $\lambda$, this 
interpretation being independent of the dimension $D$.
Our purpose is to relate the function $q(\lambda)$ with $\hat{g}(s)$ 
the Laplace
Transform of the function $g(t)$  which, as we have shown in the preceding
subsection, determines univocally the decay law.   
Employing (\ref{qt}) and (\ref{ilt}) 

\begin{equation}
\label{q} 
Q(t)= 1- \int_{^0}^{\infty} q(\lambda) \;  e^{-\lambda t} \; d\lambda
\end{equation}
and making the Laplace Transform (LT) of (\ref{q})

\begin{equation}
\label{Stiel}
 \int_{^0}^{\infty} \frac{q(\lambda)}{\lambda +s } \; d\lambda= \frac{1}{s}-
\hat{Q}(s)
\end{equation}
 
The left hand side of (\ref{Stiel}) is the Stieltjes Transform (ST)
 of the function $q(\lambda)$ and arises naturally from the iteration of the 
LT.
Inverse transforming (\ref{Stiel}) we can determine univocally
$q(\lambda)$ from $\hat{Q}(s)$.
Here we do not give the exact expression for the inversion of (\ref{Stiel}).
 For further details see \cite{widder}.
We would like to emphasize that this point of view enables us to determine the
fraction of initial conditions that decay at time t with a decay  rate 
 between $\lambda$ and $\lambda +d\lambda$ when the function $\hat{g}(s)$ 
is known.

\section{SUMMARY AND CONCLUSIONS}
\label{con}
In the present work we have studied the decay of the Sinai well in $D$
dimensions, this system being an extension to higher dimensions of 
the previous one studied in detail in I.
The main difference between the $D>2$ dimensional system and the analogue in
two dimensions is that being both  completely chaotic, the first one
has an invariant parabolic trapped set for all the values of the radii $R$
of the scatterer, whereas in the second one the invariant trapped set can be
fully hyperbolic or have a parabolic trapped subset 
if $R<\frac{\sqrt{2}}{4}$.   
In terms of the decay law it means that for the present system the decay law
is always  of algebraic type for long times (related to initial conditions
 asymptotic to the parabolic periodic orbits in $D>2$ dimensions).
We have related the decay law in $D$ dimensions to internal distributions
that characterize the dynamics and concluded that the exponent $\delta$
for the algebraic long time tail is $\delta=1$ irrespective of the dimension.
This behavior is in agreement with the results encountered in ref. 
\cite{bersh}. In that work the authors investigate in $D$ dimensions
a system like a Sinai billiard \cite{sinai} and allow  the  decay  providing
 a small window in one of the walls of the container.
They conclude that for the regular case (without a scatterer in the center)
the decay law is algebraic $(\sim 1/t)$ independent of $D$.
As the trapped set of the regular system coincides with the parabolic periodic
orbits of the Sinai well, the result is consistent with our conclusion.\\ 
We also studied the intermediate behavior of the decay law and concluded
 that
it is related to the so called hidden corridors, which asymptotic initial 
conditions contribute to the decay law with a temporal dependence of
the type $(1/t^{\mu})$, with  $2\leq\mu\leq{D}$.\\
The present work was partially supported by CONICET and Fundaci\'on 
Antorchas.
\newpage
\appendix
\section{}

In this appendix we derive the explicit expression for $\omega_{D}$, the 
transition probability from the bounded to the free region after one 
collision with the scatterer. It can be evaluated using ergodic theory
as the ratio between all orientations of momentum in the free region and all 
possible orientations of momentum.\\
The space of momenta is a $D$ dimensional unit sphere (the modulus
 of $\vec{p}$ is one).
Let  $d\Psi_{lim}(D)$ be the solid angle subtended by $\Psi_{lim}$, where
$\Psi_{lim}$ is given in expression (\ref{limpr}), and $\Omega(D)$ the total 
solid angle in $D$ dimensions.
As we have established in (\ref{prob})  
\begin{equation}
\omega_{D} = \frac{d\Psi_{lim}(D)}{\Omega(D)} \:, 
\end{equation}

so for computing $\omega_{D}$ we must have the explicit expressions for
$\Omega(D)$ and $d\Psi_{lim}(D)$.

Introducing the polar coordinates in $D$ dimensions ($D\geq{3}$)
\begin{equation}
(r,\phi,\theta_{1},......,\theta_{D-2})
\end{equation}
with $r=1$, $0<\phi< 2\pi$ and $0<\theta_{i}<\pi$,  $i=1,....D-2$ 
we can derive the expression 
\begin{equation}
\label{solid}
\Omega(D)= 2  \pi \prod_{i=1}^{D-2}\int_{0}^{\pi} \sin^{i}\theta_{i} \;
d\theta_{i} \; .
\end{equation}

Employing the well-known formula \cite{gradsh}

\begin{equation}
\int_{0}^{\pi} \sin^{i}\theta_{i} \; d\theta_{i} = \sqrt{\pi} \;
\frac{\Gamma(i+1/2)}{\Gamma(i+2/2)} 
\end{equation}

and putting it back in (\ref{solid}), we obtain
\begin{equation}
\label{isolid}
\Omega(D)=\frac{\pi^{D-2}}{\Gamma(D/2)}
\end{equation}

In order to compute $d\Psi_{lim}(D)$ we must perform the integration in 
the variables $\theta_{i}$ with $i=1,...D-3$ between $0<\theta_{i}<\pi$, and 
$0<\theta_{D-2}<\Psi_{lim}$
\begin{equation}
d\Psi_{lim}(D)=2\pi \; \prod_{i=1}^{D-3}\int_{0}^{\pi} \sin^{i}\theta_{i} \;
d\theta_{i} \int_{0}^{\Psi_{lim}} \sin^{D-2}\theta_{D-2} \; d\theta_{D-2}
\end{equation}

after a straightfoward calculation we obtain
\begin{equation}
\label{ilim}
d\Psi_{lim}(D)= \frac{2 \; (\sqrt{\pi})^{D-1} \; \Psi_{lim}^{D-1}}{(D-1)
\; \Gamma(D-1/2)} \; ,
\end{equation}  

and from (\ref{prob}), (\ref{isolid}) and (\ref{ilim}) results
\begin{equation}
\omega_{D}= \frac{2 \; \Gamma(D/2) \; \Psi_{lim}^{D-1}}{\sqrt{\pi} \; (D-1)
\; \Gamma(D-1/2)} \; .
\end{equation}

\newpage

\begin{figure}
\caption{(a)Numerical results of $G$ for $D=2,3,4$. We have
fixed $a$ (the side of the $D$ dimensional square well) as the unity 
of length and $a/\protect\sqrt{2(E+V_{0})}$ as the unity of time. 
The log-log plot shows  $G$ vs. $t$ for  radius $R=0.23$ of the central 
scatterer. (b)The long time tail together with the best fit to 
$G$ vs. $t$ that is consistent with the exponent $\delta=1$ for $D=2,3,4$.} 
\label{234gr2p3}
\end{figure}

\begin{figure}
\caption{(a)Numerical results of $G$ for $D=2,3,4$. The units are the same 
as in FIG.\protect\ref{234gr2p3}. 
The log-log plot shows  $G$ vs. $t$ for  radius $R=0.4$ of the central 
scatterer. (b)The long time tail together with the best fit to 
$G$ vs. $t$ that is consistent with the exponent
$\delta=1$ for $D=3,4$.}
\label{234gr4p0}
\end{figure}

\begin{figure}
\caption{Numerical results of $G$ for $D=4$. The units are the same 
as in FIG.\protect\ref{234gr2p3}. 
The log-log plot shows  $G$ vs. $t$ for  radius $R=0.4$ of the central 
scatterer.
The dashed curve corresponds to initial conditions in all phase space.
The solid curve results from having not populated any principal corridors.
 The 
algebraic long time tail is originated by the hidden corridors. The best
fit is consistent with an exponent $\delta=2$.}
\label{4gr4p0st}
\end{figure}

\begin{figure}
\caption{Periodic configuration of scatterers in $D=3$ dimensions, for a
constant value of the coordinate $z$. The coordinate $z$ has been chosen
in order to cut the spherical scatterers at their centers.
The principal corridor defined by  $v_{1}/v_{2}= 1$ and $v_{3}$ arbitrary
 that satisfied the condition $\mid \vec{v}\mid =1$ is shown. The width of
the corridor is $l$, and the angle $\alpha$ $(\sim l/t^{*})$ is proportional 
to the number of initial conditions that collides with some  scatterer
in a time $t>t^{*}$. }
\label{config}
\end{figure}

\end{document}